\documentclass[twoside,12pt]{article}
\usepackage{amsmath,amssymb,amsthm,latexsym}

\usepackage{stmaryrd,wasysym,upgreek,mathrsfs,dsfont}
\usepackage[english]{babel}
\usepackage{graphicx,color}
\usepackage{slashed,subfig}

\usepackage[pdftex]{hyperref}
\newcommand{\bqa}{\begin{eqnarray}}
\newcommand{\eqa}{\end{eqnarray}}
\newcommand{\be}{\begin{equation}}
\newcommand{\ee}{\end{equation}}

\newcommand{\si}{\sigma}

\theoremstyle{plain}% default
\newtheorem{thm}{Theorem}[section]

\theoremstyle{plain}
\newcommand{\beqa}{\begin{eqnarray}}
\newcommand{\eeqa}{\end{eqnarray}}
\newcommand{\bea}{\begin{eqnarray}}
\newcommand{\eea}{\end{eqnarray}}

\renewcommand{\mathbf}{\boldsymbol}

\newcommand{\R}{\mathbb{R}}

\DeclareMathOperator*{\Tr}{Tr}

\newcommand\cF{{\mathcal F}}

\newcommand\cT{{\mathcal T}}

\numberwithin{equation}{section}

\begin{document}

\title{Why Renormalizable NonCommutative\\ Quantum Field Theories?}
\author{Vincent {\sc Rivasseau}\\
Laboratoire de Physique Th\'eorique\\
 B\^at.\ 210\\
Universit\'e Paris XI\\
 F--91405 Orsay Cedex, France}
%e-mail: \texttt{rivass@th.u-psud.fr}}

\maketitle

\begin{abstract}
A new version of scale analysis and renormalization theory has been found
on the non-commutative Moyal space. Hopefully it may
connect nicely to Alain Connes's  interpretation of the standard 
model in terms of non-commutative geometry. However it is also
a non-trivial extension of this circle of ideas. It
could be useful both for physics beyond the standard model or for standard
physics in strong external field. The good news is that quantum field theory 
is better behaved on noncommutative than on ordinary space-time.
Some models at least should be completely finite, even beyond perturbation theory. 
We discuss why the $\phi^{\star 4}_4$ theory can be built with some
extensions of the traditional methods of constructive theory. 
In this way noncommutative field theory might become a possible alternative to 
supersymmetry or to string theory, whose key properties are also to tame ultraviolet divergencies.
We suggest that by gluing together many Grosse-Wulkenhaar theories at high energy
one can obtain an effective commutative field theory at lower energy.

\end{abstract}

%\tableofcontents

\section{Introduction}

I was recently proposed at the occasion of a set of lectures at the Research Institute of Mathematical 
Sciences in Kyoto to answer the question which is the title of this paper.
In fact I have neither a unique nor a precise answer to give, but rather many different partial answers.
Renormalizable noncommutative quantum field theories, hereafter called RNCQFT 
seem to me important to study because they combine many nice features
which hint to the potential solution of many different physical riddles: the Higgs mechanism, a cure for
the Landau ghost, the strange spectrum of plateaux in the quantum Hall effect, possibly the emergence 
of effective strings in confinement, even perhaps a simplified road towards quantization 
of gravity. I realize this is a good occasion to try to clarify my own ideas and to put in shape 
some written defense of this subject, based on a condensed version of 
\cite{Rivasseau:2007ab,RV1}, together with some updating of what 
happened in the last few months. But I must warn the reader that since RNCQFT is 
only three years old and fast evolving, nothing is fixed yet. This short review
might be soon outdated, some proposals might fade away, and some 
new discoveries may throw us into unexpected directions.

I'll try to answer briefly at a completely general level why the three separate elements,
renormalizability, quantum field theory and non commutative geometry (in that perhaps
slightly surprising order) appear to me as key components of theoretical physics. 
Then it should be clear why the formalism which combines these three elements together,
namely RNCQFT, should also be studied.

I'll continue with a brief review of what has been accomplished so far in the subject
and a possible preliminary classification of RNCQFT's.
I'll develop a little longer  the constructive aspects developed recently.
Finally I'll end up with a list of possible applications
and conclude with some speculations about 
how this formalism may connect to our ordinary world.

\medskip
\noindent{\bf Acknowledgments}
\medskip

I would like to thank Prof. Keiichi Ito, from Setsunan University in Osaka, and the Research Institute of Mathematical 
Physics in Kyoto for their warm hospitality and for giving me this fresh opportunity to present again in 
a slightly different form this set of ideas. My very warm thanks also extend to all the people who contributed 
to the elaboration of this material, in particular M. Disertori, R. Gurau, J. Magnen, A. Tanasa,
F. Vignes-Tourneret, J.C. Wallet and  R. Wulkenhaar. I would like also to sincerely apologize to the many people whose work in this area would be worth of citation but who have not been cited here.

\section{Why R, NC, QFT?}

\subsection{Why renormalization?}

The history of renormalization is quite an extraordinary one. In less than about half a century it 
metamorphosed  from an obscure technique to remove infinities in quantum field theory 
into a truly  ubiquitous concept, so central to physics that I do not hesitate
to put it first in this list.

The most interesting physical phenomena,
whether they pertain to quantum field theory, phase transitions, turbulence, condensed matter behavior and so on
in general occur over many scales with at least approximate scaling laws. The exponents of these laws 
usually show some beautiful universal character. The main step to understand this universality 
was made by K. Wilson when he connected two previously different areas,
renormalization in quantum field theory and the classical evolution of dynamical systems \cite{Wil}.
This uncovered the analogy between time evolution and the evolution
of effective actions under change of the observation scale.

I consider not excessive to compare the importance of renormalization in physics  to that of DNA for biology.
Indeed renormalization theory gives us the key to understand self-replication over scales. An other
comparison with biology may come from Darwinism. In physics fixed points
of the renormalization group do appear
because they are the only ones to survive renormalization group flows.
In practice Gaussian fixed points, corresponding to 
perturbatively renormalizable theories can be fully analyzed,
keeping their fundamental structure unchanged, but with a few rescaled parameters.
One should not believe that such just renormalizable theories are rare: for instance the Fermi
surface singularities make the usual Fermionic non relativistic theories of condensed matter
just renormalizable in \emph{any} space dimension \cite{FT1,FT2}.

Let us make an additional remark which points to another fundamental
similarity between renormalization group flow
and time evolution. Both seem naturally \emph{oriented} flows.
Microscopic laws are expected to determine macroscopic laws, not the converse.
Time runs from past to future and entropy increases rather than decreases.
This is philosophically at the heart of standard determinism.
A key feature of Wilson's RG is to have defined in a mathematically precise way 
\emph{which} short scale  information should be forgotten through coarse graining: it is the 
part corresponding to the \emph{irrelevant operators} in the action.
But coarse graining is also fundamental
for the second law in statistical mechanics, which is the only law
in classical physics which is ``oriented in time" and also the one 
which can be only understood in terms of change of scales.

Whether this arrow common to RG and to time evolution
is of a cosmological origin remains to be better understood. We remark simply here that
in the distant past the big bang has to be explored and understood
on a logarithmic time scale. At the beginning of our universe
important physics occurs at very short distance.
As time passes and the universe evolves, physics at longer distances,
lower temperatures and lower momenta becomes literally visible.
Hence the history of the universe itself can be 
summarized as a giant unfolding of the renormalization group.

This unfolding can then be specialized into many
different technical versions depending on the particular 
physical context, and the particular problem at hand. 

However there is one domain, namely quantum gravity, which seems difficult
to reconcile with renormalization. Ordinary quantum gravity is not renormalizable in the ordinary sense.
We learn from string theory that one should expect surprises
in the new notions of scale and renormalization group that govern 
physics at Planckian or transPlanckian energies. 
In particular the string dualities mix ordinary notions of short and 
long lengths: winding modes around short circles
cannot be distinguished from translation modes around long circles and vice versa.
This feature is also present in RNCQFT (but in a much more 
accessible mathematical setting) because  instead of being based
on the heat kernel, RNCQFT's are based on the Mehler kernel
which exhibit duality between small and large lengths.

\subsection{Why Quantum Field Theory?}

Here probably nobody should argue, so this is the easy part. 
In the strictest sense quantum field theory or QFT
provides a quantum description of particles and interactions which is 
compatible with special relativity. 
But the mathematical essence of quantum field theory is to treat 
by functional integration systems of infinitely many degrees of freedom. Therefore its formalism applies 
beyond particle physics to many non relativistic problems in statistical mechanics and
condensed matter. Even classical mechanics 
can benefit from a field theoretic reformulation. Hence QFT is a truly ubiquitous 
formalism. At the mathematical level it is much more advanced
than for instance string theory or loop quantum gravity.

Over the years relativistic QFT 
has evolved into the \emph{standard model} which explains in great detail
most experiments in particle physics and is contradicted by none.
But it suffers from at least two flaws. First it  lives on a rigid 
and flat space time background and is not yet compatible with general relativity.
Second, the standard model incorporates so many different Fermionic matter fields coupled by 
Bosonic gauge fields that it seems more some kind of new Mendeleyev table than a
fundamental theory. For these two reasons QFT and the standard model
are not supposed to remain valid without any changes until the
Planck length where gravitation should be quantized. They could 
in fact become inaccurate much before that scale.

\subsection{Why noncommutative geometry?}

In general noncommutative geometry
corresponds to an extension of the commutative algebra of functions on an ordinary manifold into
a noncommutative algebra.

As soon as ordinary coordinates functions no longer commute
a fundamental  dimensioned area appears proportional to their commutator. 
But there exists certainly a fundamental length 
in our world, namely $\ell_P = \sqrt{\hbar G /c^3}$,
obtained by combining the three fundamental constants of physics, Newton's gravitation constant $G$, Planck's
constant $\hbar$ and the speed of light $c$. Its value is about $1.6 \; 10^{-35}$ meters. Below this length 
gravity should be quantized, and the energy required for a particle to probe physics at such small distances seems to create 
a black hole horizon which prevents the very observation of this physics.
Most experts agree that this means that ordinary commutative flat space-time should 
be modified. Noncommmutative geometry seems a very natural possibility in this respect.
The fact that  black hole entropy involves the area of an horizon seems also to point to
the  Planck \emph{area} as being more fundamental than the Planck \emph{length},
just as should be the case if there is a non-trivial commutator between coordinates.

Following initial ideas of Schr\"odinger and Heisenberg  \cite{Schro,Heis} who tried to extend the 
noncommutativity of phase space to ordinary space, noncommutative 
quantum field theory was first formalized by Snyder \cite{Snyder} in the hope
that it should behave better than ordinary field theory in the ultraviolet regime.

After initial work by Michel Dubois-Violette, Richard Kerner and John Madore  \cite{DKM},
Alain Connes, Ali Chamseddine and others have forcefully advocated 
that the \emph{classical} Lagrangian of the current standard
model arises much more naturally on simple non-commutative geometries
than on ordinary commutative Minkowski space \cite{Connes:1994yd}, and leads
naturally to the classical Einstein Hilbert action for gravity.
The  noncommutative reformulation initially threw light on the Higgs mechanism
and later on more and more detailed aspects of the standard model.
We have now a fairly compelling picture: the detailed Lagrangian of the standard model
can be reproduced very simply from the principle of a spectral action corresponding
to a Dirac operator on a manifold which is ordinary space-time $R^4$ twisted
by a simple non commutative finite dimensional "internal" algebra  \cite{Barrett,Connes}.

The interest for non commutative geometry came out also from string theory.
Open string field theory may be recast as a problem of noncommutative
multiplication of string states \cite{Witten}.
String theorists realized in the late 90's that NCQFT is an effective theory of strings
\cite{a.connes98noncom,Seiberg1999vs}. Roughly this 
is because in addition to the symmetric tensor $g_{\mu\nu}$ the spectrum 
of the closed string also contains an antisymmetric tensor $B_{\mu\nu}$. There is no reason
for this antisymmetric tensor not to freeze at some lower scale into a classical field,
just as $g_{\mu\nu}$ is supposed to freeze into the classical metric of
Einstein's general relativity. But such a freeze of $B_{\mu\nu}$ 
precisely induces an effective non commutative geometry. 
In the simplest case of flat Riemannian metric and trivial constant antisymmetric tensor,
the geometry is simply of the Moyal type; it reduces to 
a constant anticommutator between (Euclidean) space-time coordinates. 
This made NCQFT popular among string theorists. A good review of these ideas 
can be found in \cite{DouNe}.

Let us remark also that NCQFT is also the right setting for down 
to earth applications such as quantum physics in strong external field
(e.g. in condensed matter the Quantum Hall effect  \cite{Suss}-\cite{Poly}-\cite{HellRaam}).

\section{RNCQFT: the present state}

NCQFT combines nicely the last  two of the three elements above.
The Connes-Chamseddine version of the standard model remains in the line of Einstein's classical unification of Maxwell's electrodynamics equations through the introduction of a new four dimensional space-time. 
Climbing in energy, the next logical step seems to find the quantum version of these ideas, 
which ought to be quantum field theory on non-commutative geometry, or NCQFT. 
We may indeed see noncommutativity, initially confined
to some internal space, invade more fully spacetime itself as we approach the Planck scale. 
Going down in energy from the Planck scale (at which at least part of string theory may be correct), 
we may also meet NCQFT's as effective theories. 
Therefore it is tempting to think that  there ought to be some intermediate regime
between QFT and possibly string theory (or another theory of quantum gravity)
where NCQFT is the right formalism. The ribbon graphs 
of NCQFT may be interpreted either as ``thicker particle world-lines"
or as ``simplified open strings world-sheets"
in which only the ends of strings appear but not yet their internal
oscillations. 

These two lines of arguments both point to develop
NCQFT.  However remember we really want in fact
RNCQFT because we argued that renormalizable theories
are the building blocks of physics, the ones who
survive RG flows. 

Until recently a big stumbling block remained on this road. 
The simplest NCQFT on Moyal space,
such as $\phi^{\star 4}_4$, were found not to be renormalizable because of a surprising phenomenon called \emph{uv/ir mixing}. The $\phi^{\star 4}_4$ theory still has infinitely many
ultraviolet divergent graphs but fewer than the ordinary $\phi^{4}_4$ theory. However ultraviolet
convergent graphs, such as the non-planar tadpole \includegraphics[scale=.7]{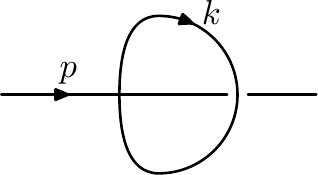} generate unexpected infrared divergences which are not of the renormalizable 
type \cite{MiRaSe}.

However three years ago the solution out of this riddle was found.
H. Grosse and R. Wulkenhaar in a brilliant series of papers
discovered how to renormalize $\phi^{\star 4}_4$ \cite{GrWu03-1,c}
on four dimensional flat non-commutative Moyal space.

The first renormalization proof \cite{c} was based on 
the matrix representation of the Moyal product. It relies on adding to 
the usual propagator a marginal harmonic potential, as required by 
Langmann-Szabo duality \cite{LaSz}. We now call such an addition
which allowed renormalization the \emph{vulcanization}\footnote{\label{vulca} 
Vulcanization is a technological operation
which adds sulphur to natural rubber to improve its mechanical properties and its resistance to
temperature change, and temperature is a scale in imaginary time...} of the model.
Vulcanization means that NCQFT on Moyal spaces has to be based on the Mehler kernel,
which governs propagation in a harmonic potential,
rather than on the heat kernel, which governs 
ordinary propagation in commutative space.
Grosse and Wulkenhaar were able to compute for the first time the Mehler kernel in the
\emph{matrix base} which transforms the Moyal product 
into a matrix product.  They combined this  computation 
with an extensive analysis of all possible 
contractions of ribbon graphs in the RG equations for the
corresponding class of so-called quasi-matrix models.
In this way they proved perturbative renormalizability of the theory, 
up to some estimates which were finally proven in \cite{Rivasseau:2005bh}. 

These founding papers opened up the subject of renormalizable non commutative field
theories, hereafter called RNCQFT.
By matching correctly propagator and interaction to respect symmetries, 
Grosse and Wulkenhaar had followed one of the main successful thread of 
quantum field theory.

The initial renormalization proof was improved by introducing
multi-scale analysis, first in the matrix base \cite{Rivasseau:2005bh},
then in position space \cite{Gurau:2005gd}. The 
$\beta$-function was computed at one loop in  \cite{Grosse:2004by}, then 
shown to vanish (first  up to three loop order and then to all orders) in \cite{Disertori:2006uy,Disertori:2006nq} 
at the self-duality point $\Omega=1$ (where $\Omega$ is the frequency of the harmonic
term). The exciting conclusion is that the $\phi^{\star 4}_4$-theory is asymptotically safe,
hence free of any Landau ghost. Wave function renormalization exactly compensates
the renormalization of the four-point function, so that the flow
between the bare and the renormalized coupling is bounded.

Essentially most of the standard tools of field theory
such as parametric \cite{Gurau:2006yc,Rivasseau:2007qx}  and Mellin representations
\cite{Gurau:2007az}, dimensional regularization and renormalization \cite{Gurau:2007fy} and
the Connes-Kreimer Hopf algebra formulation of renormalization \cite{Tanasa:2007xa}
have now been generalized to RNCQFT. Other models have been also developed 
In \cite{VignesTourneret:2006nb,VignesTourneret:2006xa} renormalization
to all orders of the duality-covariant orientable Gross-Neveu model
was shown; the one-loop beta function of the model was computed in
\cite{Lakhoua:2007ra}. The Dirac operator in
\cite{VignesTourneret:2006nb,VignesTourneret:2006xa} is \emph{not} the
square root of the harmonic oscillator Hamiltonian appearing in the
$\phi^4$-model of \cite{c} but is
of the \emph{covariant} type studied (for scalar fields) in
\cite{Langmann:2003cg,Langmann:2003if}, 
hence describes the influence
of a constant magnetic background field. Its spectrum has infinite degeneracy.
This fact can also be seen from a different structure of the
propagator in position space \cite{Gurau:2005qm}. It makes the
renormalization of the Gross-Neveu model technically more difficult,
but to understand such  \emph{covariant} models is important for the future
application of RNCQFT to condensed matter problems such as the quantum Hall effect.

Concerning other just renormalizable scalar models, in \cite{Grosse:2005ig,Grosse:2006tc} 
the noncommutative $\phi^{\star 3}_6$-model at the  self-duality point was built and 
shown just renormalizable and exactly solvable. Self-duality  relates  
this  model  to   the  Kontsevich-model.   For
$\phi^{\star 3}_4$, see \cite{Grosse:2006qv}. The $\phi^{\star 6}_3$-model 
has been shown renormalizable with $x$-space techniques  in \cite{zhituo}.
That model however is not expected to have a non-perturbative definition because
it should be unstable at large $\phi$.

A remaining very important and difficult goal is to properly "vulcanize" gauge theories
such as Yang-Mills in four dimensional Moyal space or Chern-Simons on the two
dimensional Moyal plane plus one additional ordinary commutative time direction.
We do not need to look at complicated gauge groups since the $U(1)$ pure gauge theory 
is non trivial and interacting on non commutative geometry even without matter fields. 
What is not obvious is to find a proper compromise between gauge and Langmann-Szabo 
symmetries which still has a well-defined perturbation theory  around a computable vacuum
after gauge invariance has been fixed through appropriate Faddeev-Popov or BRS procedures. 
We should judge success in my opinion by one main criterion, namely
renormalizability. Non commutative action for gauge fields which can be induced 
through integration of a scalar renormalizable matter field minimally coupled to the 
gauge field have been computed independently by de Goursac, Wallet and Wulkenhaar 
\cite{GourWW}, and by Grosse and Wohlgenannt. \cite{Grosse:2007jr}.
The result exhibits both gauge symmetry and LS covariance,
hence vulcanization, but the vacuum looked non trivial so that to check whether the associated 
perturbative expansion is really renormalizable seems difficult. Recently a new progress was accomplished
by Grosse and Wulkenhaar  \cite{Grosse:2007jy}. They showed firstly that  this vulcanized gauge action 
is the gauge part of a more general action including Higgs fields that can be deduced from
the Connes-Lott spectral action, and secondly they found the equation obeyed by a radial non-trivial vacuum. 
In this work, it appears much more clearly that the harmonic potential of
the initial Grosse Wulkenhaar model is intimately related to the symmetry breaking of
Higgs model which produces also this non-trivial gauge vacuum.

\section{A short classification of RNCQFT}

We can now propose a preliminary classification of these models into 
different categories, according to the behavior of their propagators:

\begin{itemize}
\item ordinary models at $0 < \Omega < 1$ such as $\Phi^{\star 4}_4$ 
(which has non-orientable graphs) or $(\bar\phi\phi)^2$ models 
(which has none). Their
propagator, roughly $(p^2 + \Omega^2 \tilde x^2 + A)^{-1}$ is LS covariant and has good decay both in 
matrix space and direct space. They have non-logarithmic mass 
divergencies and definitely require ``vulcanization'' i.e. the $\Omega$ term.
\item self-dual models at $\Omega = 1$ in which the propagator is LS 
invariant.
Their propagator is even better. In the matrix base 
it is diagonal, e.g. of the form $G_{m,n}= (m+n + A)^{-1}$, where
$A$ is a constant. The supermodels seem generically 
ultraviolet fixed points of the ordinary models, at which non-trivial Ward identities
force the vanishing of the beta function. The flow of $\Omega$ to the $\Omega = 1$ fixed
point is very fast (exponentially fast in RG steps).
\item covariant models such as orientable versions of LSZ or Gross-Neveu (and presumably 
orientable gauge theories
of various kind: Yang-Mills, Chern-Simons...). They may have only logarithmic divergencies and 
apparently no perturbative UV/IR mixing. However the vulcanized version 
still appears the most generic framework for their treatment.
The propagator is then roughly $(p^2 + \Omega^2 \tilde x^2 + 2\Omega \tilde x \wedge p)^{-1}$.
In matrix space this propagator shows definitely a weaker decay
than for the ordinary models, because of the presence of a non-trivial saddle point.
In direct space the propagator no longer decays with respect to the long variables, 
but only oscillates. Nevertheless the main lesson is that 
in matrix space the weaker decay can still be used; and in
$x$ space the oscillations can never be completely killed by the vertices
oscillations. Hence these models retain therefore essentially the power counting 
of the ordinary models, up to some nasty details concerning the
four-point subgraphs with two external faces.
Ultimately, thanks to a little conspiration in which the
four-point subgraphs with two external faces are renormalized by the mass renormalization,
the covariant models remain renormalizable. This is the main message of 
\cite{VignesTourneret:2006xa}.

\item self-dual covariant models which are of the previous type but at $\Omega = 1$. 
Their propagator in the 
matrix base is diagonal and depends only on one index $m$
(e.g. always the left side of the ribbon). It is of the form $G_{m,n}= (m + A)^{-1}$.
In $x$ space the propagator oscillates in a way that often
exactly compensates the vertices oscillations. These models have definitely
worse power counting than in the ordinary case, with e.g. quadratically divergent 
four point-graphs (if sharp cut-offs are used). Nevertheless Ward identities
can presumably still be used to show that they can still be renormalized. This 
probably requires a much larger conspiration to generalize
the Ward identities of the supermodels.
\end{itemize}

Notice that the status of non-orientable covariant theories is not yet clarified.

\section{Constructive NCQFT}

Constructive field theory \cite{GJ,Riv1} builds rigorously the correlation functions
for particular field theories whose Taylor expansions in the coupling
are those of ordinary perturbative field theory. Since any formal power series
is asymptotic to infinitely many smooth functions, perturbative field theory alone
does not provide any well defined recipe to compute to arbitrary accuracy
a given physical quantity, so in a deep sense it is no theory at all. Therefore for uncompromising minds,
the only meaningful quantum field models are those of constructive field theory.

In field theory infinite volume quantities
are expressed by connected functions. One main advantage of perturbative field theory 
is that connected functions are simply the sum of the connected Feynman graphs.
But the expansion diverges because there are too many such graphs.

In fact connectedness does not require the full knowledge of a Feynman graph,
with all its loop structure, but only the knowledge of a spanning tree. 
To summarize constructive theory, it is about 
working with the trees and  hiding the quantum loops. 
Hence the constructive golden rule:

\emph{``Thou shall not know the loops, or thou shall diverge!"}

Therefore tree formulas are among the main technical tools
in constructive field theory. These formulas
lie at the root of all the constructive expansions such as the cluster expansion of 
Glimm, Jaffe and Spencer \cite{GJS}. They have been improved
over the years by Brydges, Battle, Federbush, Kennedy and others. 
The final form presented below is due to \cite{AR1};
it is a canonical formula which distributes graphs according to 
underlying trees in a completely symmetric and positivity-preserving way. 

\begin{thm}
Let $F$ be a smooth function of $n(n-1)/2$ line variables $x_\ell$, $\ell = (i,j), $ $1 \le i < j \le n$.
We have
\begin{eqnarray}\label{treeformul}
F({\mathbf{1}}) = \sum_\cF \bigg\{ \prod_{\ell\in \cF}   
\big[ \int_0^1 dw_\ell \big]   \bigg\} 
 \bigg\{ \prod_{\ell\in \cF}   
\frac{\partial  }{\partial_{x_\ell} }  F  \bigg\}  \big[  x^\cF ( \{ w\}) \big] 
\end{eqnarray}
where  $F({\mathbf{1}}) $ means $F(1,1, ..., 1)$ and
\begin{itemize}
 
\item the sum over $\cF$ is over all forests over $n$ vertices,
 
\item the interpolated value $x^\cF_\ell (\{ w\})$ is 0 if $\ell = (i,j)$,
with $i$ and $j$ in different connected components with respect to $\cF$,
and is the infimum of the $w_{\ell'}$ for $\ell'$ running over the unique path from $i$ to $j'$ in $\cF$.
 
\item
Furthermore the real symmetric matrix $x^\cF_{i,j} (\{ w\})$ (completed by
1 on the diagonal $i=j$) is \emph{positive}.

\end{itemize}
\end{thm}

The constructive program launched by A. Wightman and pursued by J. Glimm, A. Jaffe
and followers in the 70's partial failed because no natural four dimensional field theory 
could be identified and fully built. This is because the only theories 
asymptotically free in the ultraviolet limit, namely non-Abelian gauge theories,
are very complicated (gauge fixing is marred by the so-called Gribov problem).
Moreover in these theories ultraviolet asymptotic freedom comes at the price of
infrared slavery: non-perturbative long range effects such as quark confinement 
are not fully understood until now, even at a non-rigorous level. 
The constructive program went on, but mostly as a set of rigorous techniques applied to many 
different areas of mathematical physics \cite{constr1,constr2}. 

For mathematical physicists who like me 
came from the constructive field theory program, the Landau ghost has always been
a big frustration.Since non Abelian gauge theories are so complicated
and lead to confinement in the infrared regime, there is no four dimensional
rigorous field theory without unnatural cutoffs up to now\footnote{We have only 
built renormalizable theories for two dimensional Fermions 
\cite{GK1}-\cite{FMRS1} and for the infrared side of $\phi^4_4$\cite{GK2}-\cite{FMRS2}.}. 
It is therefore very exciting to build 
non perturbatively the $\phi^{\star 4}_4$ theory, even if it lives on the 
unexpected Moyal space and does not obey the usual axioms of ordinary QFT.

However in order to build $\phi^{\star 4}_4$ we need to extend first in a proper way constructive methods 
This because the standard constructive cluster and Mayer expansion 
of Glimm-Jaffe-Spencer \cite{GJS} does not apply here because the interaction is non-local.
This problem can be overcome by a new expansion called the loop-vertex tree expansion 
\cite{Rivasseau:2007fr}. This expansion also solves an old problem
of ordinary constructive field theory as well, namely how to compute the thermodynamic limit 
of a $\phi^4$ theory with fixed cutoffs without using any intermediate non-canonical lattice discretization
\cite{Magnen:2007uy}. Hence it provides a first example
where the stimulation of NCQFT leads to a new tool for \emph{ordinary} field theory.

As a  first step we have proved Borel summability \cite{Sok} in the coupling constant 
for the connected functions in a way which has to be \emph{uniform} in the slice index.
The full construction of $\phi^{\star 4}_4$-theory now requires to extend these tools to
a multiscale analysis. The full control of the bounded RG trajectory
should then presumably follow as in \cite{Abd2}.

We now summarize  the loop-vertex tree expansion of
\cite{Rivasseau:2007fr} for a matrix model which
mimics a single slice of the $\phi^{\star 4}_4$
model. This matrix $\phi^4$ model is made of 
a Gaussian independent identically distributed measure on $N$ by $N$
real or complex matrices perturbed by a positive
$\frac{\lambda}{N} \Tr \phi^\star \phi \phi^\star \phi$
interaction. The $N \to \infty$ limit is given by planar graphs. It can be studied through various 
methods such as orthogonal polynomials, supersymmetric 
saddle point analysis and so on.
However none of these methods seems exactly suited to 
constructive results such as uniform Borel summability
in $N$  (Theorem \ref{borelunif} below).

Consider the complex case (the real case being similar). The normalized interacting measure is
\begin{equation} \label{functional}
d\nu (\Phi) = \frac {1}{Z(\lambda,N)}
e^{-\frac {\lambda}{N} 
\Tr \Phi^\star \Phi \Phi^\star \Phi} d\mu(\Phi) 
\end{equation}
where 
\begin{equation} d\mu = \pi^{-N^2}
e^{-\frac 12 \Tr \Phi^{\star}  \Phi } \prod_{i,j} d \Re \Phi_{ij} d \Im \Phi_{ij} 
\end{equation}
is the normalized Gaussian measure with covariance
\begin{equation}
<\Phi_{ij} \Phi_{kl}>= <\bar \Phi_{ij} \bar \Phi_{kl}>  = 0, \ \ 
<\bar \Phi_{ij} \Phi_{kl}> = \delta _{ik} \delta_{jl} .
\end{equation}

For the moment assume the coupling $\lambda$ to be real positive and small. 
We want to prove a uniform Borel summability theorem as $N \to \infty$ of the pressure
$N^{-2} \log  Z(\lambda, N)$, which is the suitably normalized sum of connected vacuum graphs.
This should at first sight be easy because the limit as $N \to \infty$ of this quantity is given by the sum of
all connected \emph{planar} vacuum graphs, hence is
\emph{analytic} in $\lambda$. But there is a subtlety: 
in the matrix base it seems that one needs to know all the loops
to find the correct scaling (a factor $N$ per vertex compensates the $1/N$
in the coupling). Indeed contrary to vector models, each propagator 
carries $two$ delta functions for matrix indices, one for the right and one for the left. 
At each vertex \emph{four} indices meet. A spanning tree 
provides only $n-1$ lines, hence $2(n-1)$ identifications. Therefore about $2n$
indices remain to be summed, hence two per vertex if we do not know the
loop structure. Of course in a vector model only \emph{two} indices meet at each vertex,
and each propagator carries one delta function, so that knowing a tree there remains
about one index to sum per vertex. This is why vector models
can be treated through cluster expansions.

The solution to this riddle is to sort of exchange the role
of vertices and propagators. We decompose the $\Phi$ functional integral according to an intermediate
Hermitian field $\sigma$ acting either on the right or on the left index. 
For instance the normalization $Z(\lambda, N)$ can be written as:
\begin{equation}\label{rightfield}
Z(\lambda, N)  = \int d\mu_{GUE}(\si^R) 
e^{- \Tr \log (1\otimes 1 + i \sqrt{\frac{\lambda}{N}} 1 \otimes \sigma^R ) }
\end{equation}
where $d\mu_{GUE}$ is the standard Gaussian measure on an Hermitian
field $\sigma^R$, that is the measure with covariance 
$<\sigma^R_{ij} \sigma^R_{kl}>= \delta _{il} \delta_{jk}$.  It is convenient to view  $\mathbb{R}^{N^2}$ 
as $\mathbb{R}^{N}\otimes \mathbb{R}^{N}$.
For instance the operator $H=\sqrt{\frac{\lambda}{N}}[1 \otimes \sigma^R ]   $
transforms the vector $e_{m}\otimes e_{n}$ into $\sqrt{\frac{\lambda}{N}}e_{m} \otimes \sum_k \sigma^R_{kn} e_k$.
Remark that this is a Hermitian operator because $\sigma^R$ is Hermitian.
The $e^{-\Tr\log }$ 
represents the Gaussian integration over $\Phi$, hence a big $N^2$ by $N^2$
determinant.

By duality of the matrix vertex, there is an exactly similar formula 
but with a left Hermitian field $\sigma^L$ acting on the left index, 
and with  $[\sigma^L \otimes 1 ]$ replacing $ [1 \otimes \sigma^R ]$.
From now on we work only with the right field and drop the $R$ superscript for simplicity.

We define the  \emph{loop vertex} $V$  to be
\begin{equation}\label{vertex}
V=- \Tr \log (1\otimes 1 + 1 \otimes i H ) ,
\end{equation}
and expand the exponential in (\ref{rightfield})
as $\sum_n \frac{V^n}{n!} $. To compute the connected
graphs we give a (fictitious) index $v=1,..., n$ to all the $\sigma$ fields of a given 
loop vertex $V_v$. At any order $n$ the functional integral over $d\nu (\sigma)$ 
is obviously also equal to the same integral but with a Gaussian measure $d\nu (\{\sigma^v\})$ with degenerate covariance $<\sigma^v_{ij} \sigma^{v'}_{kl}>= \delta _{il} \delta_{jk}$.

Then we apply the tree formula and we get
\begin{thm}
\begin{eqnarray}\label{treeformula}
log Z(\lambda, N) = 
\sum_{n=1}^{\infty}   \sum_\cT 
\bigg\{ \prod_{\ell\in \cT}  
\big[ \int_0^1 dw_\ell \sum_{i_\ell, j_\ell, k_\ell, l_\ell} \big]\bigg\} 
\int  d\nu_\cT (\{\sigma^v\}, \{ w \})  \nonumber \\
 \bigg\{ \prod_{\ell\in T} \big[ \delta _{i_\ell l_\ell} \delta_{j_\ell k_\ell}
 \frac{\delta}{\delta \sigma^{v(\ell)}_{i_\ell, j_\ell}}
 \frac{\delta}{\delta \sigma^{v'(\ell)}_{k_\ell, l_\ell}} 
\big] \bigg\} \prod_v V_v 
\end{eqnarray}
\end{thm}

In this way we have an expansion whose tree lines are intermediate field propagators.
No wonder they were not seen in the standard cluster expansion,
because these lines come from the former vertices of the ordinary theory!
See fro instance as an example of a particular tree on loop vertices:
\vskip 1cm
{
{\centerline{\includegraphics[scale=.3,angle=90]{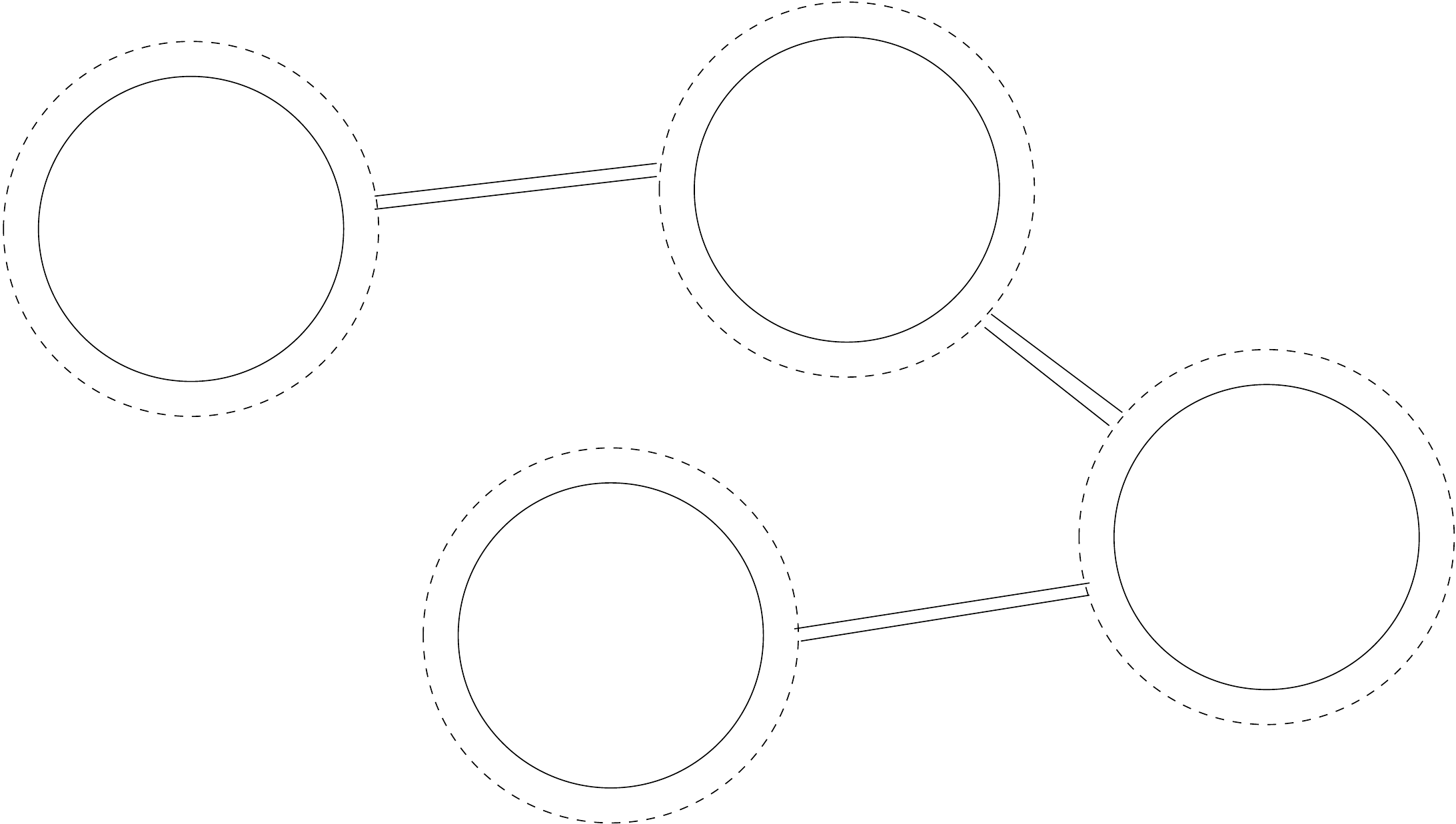}}}
}

In this way we can prove:

  \begin{thm}\label{borelunif}
The series (\ref{treeformula}) is absolutely convergent for $\lambda$ small enough
and Borel summable \emph{uniformly} in $N$.
\end{thm}

\noindent{\bf Proof:}\   Left indices provide a particular cyclic order at each loop vertex.
The $\sigma $ field acts only on right indices, hence left indices are conserved, and there is a single
global $N$ factor per loop vertex  coming from the trace over the left index. But there is a single trace over right indices corresponding to turning around the tree 
with of a product of resolvents bounded by 1!

We have started already to work on the multiscale version of this result, which is a bit more complicated.
The naive idea is that one should now optimize the expansion by building the tree in (\ref{treeformula})
in priority between  "high momentum" loop vertices, for instance whose
with highest values of the left index.
But because in the matrix base for $\phi^{\star 4}$ the same $\sigma$ field with 
low indices (for instance $\sigma_{00}$) in present both in "high momentum" 
and "low momentum" loop vertices, it seems at first sight that this 
"optimization of the tree over scales" cannot be done in a positivity preserving way.
Fortunately it is not necessary to preserve the imaginary character of all
the fields in loop vertices, when they are smeared against
a positive Gaussian measure. The corresponding formulas will be definitely more
complicated than in a single slice, but preliminary results
\cite{MR2} indicate that the whole construction can be made
without any cluster or Mayer expansion at any stage.

Let us add some words about the way in which ordinary constructive field theory
can be renewed by these ideas. Using the "loop vertex" tree expansion 
in the commutative context it is presumably possible to 
repeat classical constructions such as those of the 
infrared limit of ordinary critical $\phi^4_4$ theory
\emph{ without} using any discretization, lattice of cubes, cluster and
Mayer expansions at any stage.

This was done for a single scale model in \cite{Magnen:2007uy},
where we have shown that integration by parts combined wit the loop vertex expansion
can prove the scaled decrease of the correlation functions, through a Fredholm type inequality.

It is an interesting non-trivial problem to generalize this intermediate field method 
to higher order interactions than $\phi^4$, for instance
$\phi^{2n}$. More intermediate fields are obviously required.

Let us remark that we also expect many applications of this new method
to constructive gluing of different expansions in ordinary QFT.

\section{RNCQFT: possible applications} 

We would like now to comment further on possible areas of physical applications
of RNCQFT:

\begin{itemize}

\item {\bf  Quantum Hall Effect}

NCQFT and in particular the non commutative Chern Simons theory has been 
recognized as effective theory of the quantum Hall effect
already for some time \cite{Suss}-\cite{Poly}-\cite{HellRaam}.
But the discovery of the vulcanized RG holds promises for a better explanation
of how these effective actions are generated from the microscopic level.

In this case there is an interesting reversal of
the initial GW (Grosse-Wulkenhaar) problematic. In the
$\phi^{\star 4}_4$ theory the vertex is given
a priori by the Moyal structure, and it is LS invariant. The challenge was
to find the right propagator  which makes the theory renormalizable, and it turned out to
have LS duality. 

Now to explain the (fractional) quantum Hall effect, which is a bulk effect
whose understanding requires electron interactions, we can almost invert this logic. The propagator
is known since it corresponds to non-relativistic electrons in two dimensions
in a constant magnetic field. It has LS duality. But the effective 
theory should be anionic hence not local. 
Here again we can argue that among all possible non-local interactions, a few renormalization
group steps should select the only ones which form a renormalizable theory with the corresponding
propagator. In the commutative case (i.e. zero magnetic field)
local interactions such as those of the Hubbard model are just 
renormalizable in any dimension because of the extended nature of the Fermi-surface singularity.
Since the non-commutative electron propagator (i.e. in non zero magnetic field)
looks very similar to the GW propagator (it is in fact a 
generalization of the Langmann-Szabo-Zarembo propagator)
we can conjecture that the renormalizable interaction corresponding to this propagator
should be given by a Moyal product. That's why we hope
that non-commutative field theory and a suitable generalization of the Grosse-Wulkenhaar renormalization group
might be the correct framework for a 
microscopic {\it ab initio} understanding 
of the fractional quantum Hall effect which is currently lacking.

\item {\bf Charged Polymers in Magnetic Field}

Just like the heat kernel governs random motion,
the covariant Mahler kernel should govern random motion of charged particles 
in presence of a magnetic field. Ordinary polymers can be studied as random walk
with a local self repelling or self avoiding interaction. They can be treated by 
QFT techniques using the $N=0$ component limit or the supersymmetry trick
to erase the unwanted vacuum graphs. Many results, such as various exact 
critical exponents in two dimensions, approximate ones in three dimensions,
and infrared asymptotic freedom in four dimensions have been computed
for self-avoiding polymers through renormalization group techniques.
In the same way we expect that charged polymers under magnetic field might be studied
through the new non commutative vulcanized RG. The relevant interactions
again should be of the Moyal rather than of the local type, and there is no reason that
the replica trick could not be extended in this context. Ordinary observables such as
$N$ point functions would be only translation \emph{covariant}, but translation invariant physical
observables such as density-density correlations
should be recovered out of gauge invariant observables. In this way it 
might be possible to deduce new scaling properties of 
these systems and their exact critical exponents through the generalizations of the
techniques used in the ordinary commutative case \cite{Duplantier}.

More generally we hope that the conformal invariant two dimensional theories, the RG flows between them and the $c$ theorem of Zamolodchikov \cite{Zam} might have 
appropriate \emph{magnetic} generalizations  involving vulcanized flows
and Moyal interactions.

\item
{\bf Quark Confinement}

Quark confinement corresponds to
a strong coupling non-perturbative regime of 
non-Abelian gauge theory on ordinary commutative space.
In \cite{Seiberg1999vs} a mapping is proposed
between ordinary and non-commutative gauge fields which do not preserve the gauge groups
but preserve the gauge equivalent classes. 
The effective physics of confinement should be
governed by a non-local interaction, as is the case in effective strings or bags models.
In the initial matrix model approach of 'tHooft \cite{thooft} to this problem,
the planar graphs dominate
because a gauge group $SU(N)$ with $N$ large.
But the planar limit emerges more naturally out of NCQFT since it 
is then a renormalization group effect. The
large $N$ \emph{matrix} limit in NCQFT's  parallels the large 
$N$ \emph{vector} limit which allows to understand the formation of
Cooper pairs in supraconductivity \cite{FMRT1}. In that case $N$ is not arbitrary but
is roughly the number of effective quasi particles or sectors around the extended Fermi 
surface singularity at the superconducting transition temperature.
We called this phenomenon a \emph{dynamical} large $N$ \emph{vector} limit.
RNCQFT's provides us with the first clear example of a 
\emph{dynamical} large $N$ \emph{matrix} limit. We hope therefore that it should be ultimately useful 
to understand bound states in ordinary commutative non-Abelian gauge theories, hence quark confinement.

\item
{\bf Quantum Gravity}

Although ordinary renormalizable QFT's now seem to have 
renormalizable analogs on the Moyal space, there is no renormalizable commutative field theory
for spin 2 particles, so that the RNCQFT alone should 
not allow quantization of gravity. However
quantum gravity might enter the picture at a later and more advanced stage.
The two current main tentatives to quantize gravity 
are string theory and loop gravity\footnote{There is also the intriguing possibility that the
ordinary Einstein-Hilbert theory might have
a non trivial renormalization group fixed point, so might be renormalizable
in a non-perturbative sense \cite{Lit}.}.
We  remarked already that NCQFT 
appears as some effective version of string theory.
But ribbon graphs have borders hence corresponds to 
open strings world sheets, whenever  gravity occurs in the closed strings sector. Therefore it may have
to do with doubling the ribbons of some NCQFT 
in an appropriate way. Because there is no reason
not to quantize the antisymmetric tensor $B$ which defines the non commutative geometry
as well as the symmetric one $g$ which defines the metric, we should clearly no longer 
limit ourselves to Moyal spaces. A first step towards a non-commutative
approach to quantum gravity might be to search for
the proper analog of vulcanization in more general non-commutative 
geometries such as solvable symmetric spaces \cite{Bieliavsky1}. 

The loop gravity approach is based on a background invariant 
formulation in which a huge symmetry group, those of diffeomorphisms 
is quotiented out. It seems at first sight farther from NCQFT. But 
some contact may appear when we better understand the role of new symmetries
in RNCQFT, such as the LS duality. 

However we have to admit that any theory of quantum gravity will probably remain highly 
conjectural for many decades or even centuries.

\end{itemize}

\section{How to recover the ordinary world?}

If at some energy scale in the \emph{terra incognita} that lies 
between  the Tev and the Planck scale noncommutativity escapes
some internal space of the Connes-Chamseddine type and invades  
ordinary space-time itself, it might manifest itself first in the form
of a tiny non-zero commutator between pairs of space time variables.
From that scale up, we should use the non-commutative scale decomposition and the non-commutative renormalization group rather than the ordinary one. Although we don't know yet
how to build renormalizable non-commutative gauge theories, we may
hope that  the flow corresponding to QED (which like $\phi^4_4$
suffered from the Landau ghost in the ordinary commutative world) 
should become milder and may grind to a halt in the non-commutative world.

Noncommutative models with harmonic potential and non zero $\theta$ break both 
Lorentz and translation invariance. If renormalizable 
noncommutative gauge theories are built out of some Dirac analog
of the Grosse-Wulkenhaar propagator with harmonic potential,
as is envisioned in \cite{Grosse:2007jy}, they will also break translation invariance.
 Hence if such models have anything to do with physics beyond the standard model,
one should explain how they can connect to our
ordinary Lorentz and translation invariant commutative world.

My initial impression was that perhaps only 
models of the covariant type can make a connection to such ordinary physics,
because such covariant models do not really break translation invariance
for gauge invariant physical quantities  \cite{Rivasseau:2007ab}.

But covariant models are much more singular, and in particular they do
not seem to have $\Omega =1$ fixed points of the asymptotically safe type. 
Therefore I would like to propose an other 
possible scenario, which will be developed in a future joint publication 
with R. Gurau and A. Tanasa \cite{GRT}.

In this scenario ordinary fields at lower energies do not emerge
from a single confined model of the GW type but from the zero modes of a whole
bunch of such models, which should be glued together in a coherent way.
We know that although a lattice breaks rotation invariance, the long distance effective
theory can be rotation-invariant. Furthermore the Laplacian can emerge naturally
for instance from standard nearest-neighbor ferromagnetic coupling, but also from other
generic types of short range couplings.

We therefore consider a regular (or eventually random) four dimensional lattice $\Lambda$.
To each lattice site or cell would be associated a different copy of the
GW model and these copies would be independent of each other except for their zero modes
or perhaps for a few low values of the modes in the matrix base.
Each GW model would have its own confining harmonic potential 
roughly centered around the center of the cell.
Each would exhibit a fixed point in the ultraviolet regime. In the infrared
regime the zero modes of these GW models 
would form the degrees of freedom of the ordinary commutative field theories of our world
and govern long range physics.

Such a model at first sight resembles a field theory with a naive lattice cutoff. It has
a particular scale $\Lambda_{\theta}$ (which may or may not be the Planck scale)
essentially  given by the $\theta$ parameter which would give the area
or 4d volume of the elementary lattice cells. But it has several advantages
over a naive cutoff.

\begin{itemize}

\item There would be no true cutoff in energy. Physics would not stop at scale 
$\Lambda_{\theta}$. As one climbs in energy in our commutative world,
for instance using more and more powerful colliders,
I do not see how to avoid focusing on tinier and tinier regions of space-time.
From scale $\Lambda_{\theta}$ on, one would enter into a particular "GW worldlet" 
corresponding to the inside of a given cell.
This "worldlet" can have no ultraviolet cutoff
and remain mathematically consistent up to infinite energy.

\item The bare coupling for the commutative world, hence at scale  $\Lambda_{\theta}$ 
is also the renormalized coupling for the GW worldlets. It would be the interaction
corresponding to the GW zero mode, hence form our world lower scales
it would appear local. It corresponds to the renormalizable interactions
of ordinary field theory, since they appear to correspond also to the renormalizable ones for the
GW worldlets.

\item  We would like to investigate whether the noncommutativity of space time which killed
the Landau ghost could be a substitute for supersymmetry
to tame ultraviolet flows, but without introducing new particles.
Supersymmetry tames ultraviolet flows by adding
loops of superpartners to the ordinary loops. One of the  main arguments for supersymmetry is
that it makes the three flows of the standard model 
$U(1)$, $SU(2)$ and $SU(3)$  couplings better
converge at a single unification scale (see \cite{BJLL} 
and references therein for a discussion of this subtle question).
Replacing commutative flows by noncommutative flows
at some scale before that unification scale might also do the job.

\end{itemize}

However this scenario should be much elaborated
if it is ever to become a credible alternative to supersymmetry.
In particular discovering some natural way to glue 
the "GW wordlets" seems necessary in order to develop 
the model further. A proposal will be given in \cite{GRT} but at the moment 
it is neither canonical nor unique.

 %(see rough figure below).

%\medskip
%{\hskip1.5cm{\includegraphics[scale=.4]{flowwithoutsusy.jpg}}\hskip1cm
%{\includegraphics[scale=.36]{flowwithsusy.jpg}}
%\medskip

% \bibliographystyle{utphys}
% \bibliography{biblio-articles,biblio-books}

\begin{thebibliography}{99}


\bibitem{Rivasseau:2007ab}  V.~Rivasseau, Noncommutative renormalization,
in {\it Quantum Spaces},
Progress in Mathematical Physics {\bf 53}, Birkh\"auser, (2007), arXiv:0705.0705,

\bibitem{RV1} V. Rivasseau and F. Vignes-Tourneret,
Noncommutative Renormalization,
in {\it Rigorous Quantum Field Theory}, a Festschrift for Jacques Bros,
Birkh\"auser Progress in Mathematics {\bf  251}, (2007), {{\tt hep-th/0409312}}.

\bibitem{Wil} K. Wilson, { Renormalization group and 
critical phenomena, 
II Phase space cell analysis of critical behavior},
{\it Phys. Rev. B} {\bf 4},  (1974), 3184.


\bibitem{FT1}  J. Feldman and E. Trubowitz, 
{ Perturbation theory for Many Fermion Systems}, Helv. Phys. Acta {\bf 63}
156, (1991) .
\vskip.1cm

\bibitem{FT2} J. Feldman and E. Trubowitz, { The flow of an Electron-Phonon
System to the Superconducting State}, Helv. Phys. Acta {\bf 64}
213, (1991).



\bibitem{Schro} E. Schr\"odinger, { \"Uber die Unanwendbarkeit des Geometrie im Kleinen}
Naturwiis. {\bf 31}, (1934), 342.

\bibitem{Heis} W. Heisenberg, { Die Grenzen des Anwendbarkeit des bisherigen Quantentheorie}
Z. Phys. {\bf 110} (1938), 251.

\bibitem{Snyder} H. S. Snyder, { Deformation quantization for actions of the affine group}
Phys Rev {\bf 71} (1947), 38.

\bibitem{DKM}
M. Dubois-Violette, R. Kerner and J. Madore,
{ Noncommutative differential geometry and new models of gauge theory}
J. Math. Phys. {\bf 31}, (1990), 323-330.


\bibitem{Connes:1994yd}
A. Connes, {\it Noncommutative geometry}, Academic Press Inc., San
Diego (1994) \\
available at \url{http://www.alainconnes.org/downloads.html}.


\bibitem{Barrett}
J. W. Barrett, 
{ A Lorentzian version of the non-commutative geometry of the standard model of particle physics},
arXiv:hep-th/0608221


\bibitem{Connes}
A. Connes, { Non-commutative geometry and  the standard model with neutrino
mixing}, arXiv:hep-th/0608226.


\bibitem{Witten}
E. Witten, { Non-commutative geometry and string field theory}
Nuclear Physics B, {\bf 268}, (1986),  253-294.

\bibitem{a.connes98noncom}
A.~Connes, M.~R. Douglas, and A.~Schwarz, { Noncommutative geometry and matrix
  theory: Compactification on tori}, {JHEP} {\bf 02} (1998) 003,
  \href{http://www.arXiv.org/abs/hep-th/9711162}{{\tt hep-th/9711162}}.

\bibitem{Seiberg1999vs}
N.~Seiberg and E.~Witten, { String theory and noncommutative geometry}, {
  JHEP} {\bf 09} (1999) 032,
\href{http://www.arXiv.org/abs/hep-th/9908142}{{\tt hep-th/9908142}}.
%%CITATION = HEP-TH 9908142;%%.

\bibitem{DouNe}
M.~R. Douglas and N.~A. Nekrasov, { Noncommutative field theory}, { Rev.
  Mod. Phys.} {\bf 73}, (2001), 977--1029,
\href{http://www.arXiv.org/abs/hep-th/0106048}{{\tt hep-th/0106048}}.
%%CITATION = HEP-TH 0106048;%%.


\bibitem{Suss} L. Susskind, The Quantum Hall Fluid and Non-Commutative Chern Simons Theory,
hep-th/0101029

\bibitem{Poly} A. Polychronakos, 
{ Quantum Hall states as matrix Chern-Simons theory},
JHEP 0104 (2001), 011, 
\href{http://www.arXiv.org/abs/hep-th/0103013}{{\tt hep-th/0103013}}

\bibitem{HellRaam} S. Hellerman and M. Van Raamsdonk, 
{ Quantum Hall Physics = Noncommutative Field Theory},
JHEP 0110 (2001) 039, \href{http://www.arXiv.org/abs/hep-th/0103179}{{\tt hep-th/
0103179}}


\bibitem{MiRaSe}
S.~Minwalla, M.~Van~Raamsdonk, and N.~Seiberg, { Noncommutative perturbative
  dynamics}, {\em JHEP} {\bf 02} (2000) 020,
\href{http://www.arXiv.org/abs/hep-th/9912072}{{\tt hep-th/9912072}}.
%%CITATION = HEP-TH 9912072;%%.

\bibitem{GrWu03-1}
H.~Grosse and R.~Wulkenhaar, { Power-counting theorem for non-local matrix
  models and renormalization}, {\em Commun. Math. Phys.} {\bf 254} (2005),
  no.~1, 91--127,
\href{http://www.arXiv.org/abs/hep-th/0305066}{{\tt hep-th/0305066}}.
%%CITATION = HEP-TH 0305066;%%.


\bibitem{c}
H.~Grosse and R.~Wulkenhaar, { Renormalization of $\phi^{4}$-theory on
  noncommutative ${\R}^4$ in the matrix base}, {\em Commun. Math. Phys.} {\bf
  256} (2005), no.~2, 305--374,
\href{http://www.arXiv.org/abs/hep-th/0401128}{{\tt hep-th/0401128}}.
%%CITATION = HEP-TH 0401128;%%.

\bibitem{LaSz}
E.~Langmann and R.~J. Szabo, { Duality in scalar field theory on noncommutative
  phase spaces}, {\em Phys. Lett.} {\bf B533} (2002) 168--177,
\href{http://www.arXiv.org/abs/hep-th/0202039}{{\tt hep-th/0202039}}.
%%CITATION = HEP-TH 0202039;%%.

\bibitem{Rivasseau:2005bh}
  V.~Rivasseau, F.~Vignes-Tourneret and R.~Wulkenhaar,
{ Renormalization of noncommutative $\phi^4$-theory by multi-scale  analysis},
  Commun.\ Math.\ Phys.\  {\bf 262} (2006) 565,
  [arXiv:hep-th/0501036].
  %%CITATION = CMPHA,262,565;%%

%\cite{Gurau:2005gd}
\bibitem{Gurau:2005gd}
  R.~Gurau, J.~Magnen, V.~Rivasseau and F.~Vignes-Tourneret,
{ Renormalization of non-commutative $\phi^4_4$ field theory in $x$-space},
  Commun.\ Math.\ Phys.\  {\bf 267} (2006) 515,
 [arXiv:hep-th/0512271].
  %%CITATION = CMPHA,267,515;%%

%\cite{Grosse:2004by}
\bibitem{Grosse:2004by}
  H.~Grosse and R.~Wulkenhaar,
{ The beta-function in duality-covariant noncommutative $\phi^4$-theory},
  Eur.\ Phys.\ J.\  C {\bf 35} (2004) 277,
  [arXiv:hep-th/0402093].
  %%CITATION = EPHJA,C35,277;%%

%\cite{Disertori:2006uy}
\bibitem{Disertori:2006uy}
  M.~Disertori and V.~Rivasseau,
{ Two and three loops beta function of non commutative $\phi^4_4$ theory},
  Eur.\ Phys.\ J.\  C {\bf 50} (2007) 661
  [arXiv:hep-th/0610224].
  %%CITATION = EPHJA,C50,661;%%

%\cite{Disertori:2006nq}
\bibitem{Disertori:2006nq}
  M.~Disertori, R.~Gurau, J.~Magnen and V.~Rivasseau,
{ Vanishing of beta function of non commutative $\phi^4_4$ theory to all
orders},
  Phys.\ Lett.\  B {\bf 649} (2007) 95
  [arXiv:hep-th/0612251].
  %%CITATION = PHLTA,B649,95;%%

\bibitem{Gurau:2006yc}
  R.~Gurau and V.~Rivasseau,
``Parametric representation of noncommutative field theory,''
  Commun.\ Math.\ Phys.\  {\bf 272} (2007) 811
  [arXiv:math-ph/0606030].
  %%CITATION = CMPHA,272,811;%%

%\cite{Rivasseau:2007qx}
\bibitem{Rivasseau:2007qx}
  V.~Rivasseau and A.~Tanasa,
``Parametric representation of `critical' noncommutative QFT models,''
  arXiv:math-ph/0701034.
  %%CITATION = MATH-PH/0701034;%%


\bibitem{Gurau:2007az}
  R.~Gurau, A.~P.~C.~Malbouisson, V.~Rivasseau and A.~Tanasa,
 ``Non-Commutative Complete Mellin Representation for Feynman Amplitudes,''
  Lett.\ Math.\ Phys.\  {\bf 81}, 161 (2007)
  [arXiv:0705.3437 [math-ph]].

\bibitem{Gurau:2007fy}
  R.~Gurau and A.~Tanasa,
 ``Dimensional regularization and renormalization of non-commutative QFT,''
  arXiv:0706.1147 [math-ph].

\bibitem{Tanasa:2007xa}
  A.~Tanasa and F.~Vignes-Tourneret,
 ``Hopf algebra of non-commutative field theory,''
  arXiv:0707.4143 [math-ph].



%\cite{VignesTourneret:2006nb}
\bibitem{VignesTourneret:2006nb}
  F.~Vignes-Tourneret,
``Renormalization of the orientable non-commutative Gross-Neveu model,''
  Annales Henri Poincar\'e {\bf 8} (2007) 427
  [arXiv:math-ph/0606069].
  %%CITATION = AHPJF,8,427;%%

%\cite{VignesTourneret:2006xa}
\bibitem{VignesTourneret:2006xa}
  F.~Vignes-Tourneret,
`` Renormalisation des th\'eories de champs non commutatives,'' Ph.D.\
thesis,   arXiv:math-ph/0612014.
  %%CITATION = MATH-PH/0612014;%%


%\cite{Lakhoua:2007ra}
\bibitem{Lakhoua:2007ra}
  A.~Lakhoua, F.~Vignes-Tourneret and J.~C.~Wallet,
``One-loop beta functions for the orientable non-commutative Gross-Neveu
model,''
  arXiv:hep-th/0701170.
  %%CITATION = HEP-TH/0701170;%%


%\cite{Langmann:2003cg}
\bibitem{Langmann:2003cg}
  E.~Langmann, R.~J.~Szabo and K.~Zarembo,
``Exact solution of noncommutative field theory in background magnetic
fields,''
  Phys.\ Lett.\  B {\bf 569} (2003) 95
  [arXiv:hep-th/0303082].
  %%CITATION = PHLTA,B569,95;%%

%\cite{Langmann:2003if}
\bibitem{Langmann:2003if}
  E.~Langmann, R.~J.~Szabo and K.~Zarembo,
``Exact solution of quantum field theory on noncommutative phase spaces,''
  JHEP {\bf 0401} (2004) 017
  [arXiv:hep-th/0308043].
  %%CITATION = JHEPA,0401,017;%%


%\cite{Gurau:2005qm}
\bibitem{Gurau:2005qm}
  R.~Gurau, V.~Rivasseau and F.~Vignes-Tourneret,
``Propagators for noncommutative field theories,''
  Annales Henri Poincar\'e {\bf 7} (2006) 1601
  [arXiv:hep-th/0512071].
  %%CITATION = AHPJF,7,1601;%%





\bibitem{Grosse:2005ig}
  H.~Grosse and H.~Steinacker,
``Renormalization of the noncommutative $\phi^3$-model through the  Kontsevich
model,''
  Nucl.\ Phys.\  B {\bf 746} (2006) 202
  [arXiv:hep-th/0512203].
  %%CITATION = NUPHA,B746,202;%%

%\cite{Grosse:2006tc}
\bibitem{Grosse:2006tc}
  H.~Grosse and H.~Steinacker,
``Exact renormalization of a noncommutative $\phi^3$ model in 6 dimensions,''
  arXiv:hep-th/0607235.
  %%CITATION = HEP-TH/0607235;%%

%\cite{Grosse:2006qv}
\bibitem{Grosse:2006qv}
  H.~Grosse and H.~Steinacker,
``A nontrivial solvable noncommutative $\phi^3$ model in 4 dimensions,''
  JHEP {\bf 0608} (2006) 008
  [arXiv:hep-th/0603052].
  %%CITATION = JHEPA,0608,008;%%

\bibitem{zhituo}
Z. Wang and S. Wan,
Renormalization of Orientable Non-Commutative Complex $\Phi^6_3$ Field
Theory in $x$ Space, hep-th.wzht.23443,
to appear in Ann. Henri Poincar\'e.

\bibitem{GourWW} Axel de Goursac, J.C. Wallet and R. Wulkenhaar,
Non Commutative Induced Gauge Theory
\href{http://www.arXiv.org/abs/hep-th/0703075}{{\tt hep-th/0703075}}.


\bibitem{Grosse:2007jr}
  H.~Grosse and M.~Wohlgenannt,
 ``Renormalization and Induced Gauge Action on a Noncommutative Space,''
  arXiv:0706.2167 [hep-th].

\bibitem{Grosse:2007jy}
  H.~Grosse and R.~Wulkenhaar,
 "8D-spectral triple on 4D-Moyal space and the vacuum of noncommutative gauge
 theory,''
  arXiv:0709.0095 [hep-th].



\bibitem{GJ} J. Glimm and A. Jaffe,
Quantum physics. A functional integral point of view, Springer, 2nd edition (1987).

\bibitem{Riv1} V. Rivasseau, From perturbative to constructive renormalization,
Princeton University Press (1991).


\bibitem{GJS} J. Glimm  A. Jaffe and T. Spencer,
Convergent expansion about mean field theory II. Convergence of the expansion
Ann. Phys. {\bf 101} 631-669 (1976)


\bibitem{AR1} A. Abdesselam and V.  Rivasseau, Trees, forests and jungles: a
botanical garden for cluster expansions, in Constructive Physics, ed by
V. Rivasseau, Lecture Notes in Physics 446, Springer Verlag, 1995.

\bibitem{constr1} Constructive Physics, Proceedings of the International 
Workshop at Ecole Polytechnique, Palaiseau, July 1994, 
ed by V. Rivasseau, Lecture Notes in Physics 446, Springer Verlag (1995).

\bibitem{constr2} Constructive Field Theory and Applications: Perspectives and 
Open Problems, Journ. Math. Phys. 41, 3764 (2000).

\bibitem{GK1} 
K. Gawedzki and A. Kupiainen, Gross-Neveu model through  
convergent perturbation expansions, {\it Comm. 
Math. Phys.} {\bf 102}, 1 (1985).

\bibitem{FMRS1} J. Feldman, J. Magnen, V. Rivasseau and R.  
S{\'e}n{\'e}or, A renormalizable field theory: 
the massive Gross-Neveu model in two  
dimensions, {\it Comm. Math. Phys.} {\bf 103}, 67 (1986). 

\bibitem{GK2} K. Gawedzki
and A. Kupiainen, Massless $\phi^{4}_{4}$ theory: Rigorous control
of a renormalizable asymptotically free model,
{\it Comm. Math. Phys.}  {\bf 99}, 197 (1985).

\bibitem{FMRS2} J. Feldman, J. Magnen, V. Rivasseau
and R. S{\'e}n{\'e}or, Construction of infrared $\phi^4_{4}$
by a phase space expansion, {\it Comm. Math. Phys.}  {\bf 109}, 437 (1987).


\bibitem{Sok} A. Sokal, An improvement of Watson's theorem on Borel  
summability, {\it Journ. Math. Phys}. {\bf 21}, 261 (1980).

%\cite{Rivasseau:2007fr}
\bibitem{Rivasseau:2007fr}
  V.~Rivasseau,
``Constructive Matrix Theory,''
  arXiv:0706.1224 [hep-th].
  %%CITATION = ARXIV:0706.1224;%%

%\cite{Magnen:2007uy}
\bibitem{Magnen:2007uy}
  J.~Magnen and V.~Rivasseau,
``Constructive $\phi^4$ field theory without tears,''
  arXiv:0706.2457 [math-ph].
  %%CITATION = ARXIV:0706.2457;%%

\bibitem{Abd2} A. Abdesselam,A Complete Renormalization Group Trajectory Between Two Fixed Points,
\href{http://www.arXiv.org/abs/math-ph/0610018}{{\tt math-ph/0610018}}.


\bibitem{MR2} J. Magnen and V. Rivasseau,
work in progress.


\bibitem{Duplantier} B. Duplantier,
Conformal Random Geometry in Les Houches, Session LXXXIII, 2005, Mathematical Statistical Physics, A. Bovier, F. Dunlop, F. den Hollander, A. van Enter and J. Dalibard, eds., pp. 101-217, Elsevier B. V. (2006), \href{http://www.arXiv.org/abs/math-ph/0608053}{{\tt math-ph/0608053}}

\bibitem{Zam} A.B Zamolodchikov, JETP Letters {\bf 43}, 731 (1986)

\bibitem{thooft} G. 't Hooft, A planar diagram theory for strong interactions
Nuclear Physics B, {\bf 72}, 461 (1974)

\bibitem{FMRT1} J. Feldman, J. Magnen, V. Rivasseau and E. Trubowitz,
An Intrinsic 1/N Expansion for Many Fermion Systems,
{\it Europhys. Letters}  {\bf 24}, 437 (1993).

\bibitem{Lit} 
D. Litim,  Fixed Points of Quantum Gravity,
Phys. Rev. Lett. {\bf 92}, (2004)  201301.

\bibitem{Bieliavsky1} 
Pierre Bieliavsky,
Strict Quantization of Solvable Symmetric Spaces,
J. Symplectic Geom. Volume 1, Number 2 (2002), 269-320. 


\bibitem{GRT}
R. Gurau, V. Rivasseau and A. Tanasa, in preparation.


\bibitem{BJLL} V. Barger, Jing Jiang, Paul Langacker and Tianjun Li,
{ Gauge Coupling Unification in the Standard Model},
Phys.Lett. {\bf B624} (2005) 233-238, arXiv:hep-ph/0503226


\end{thebibliography}

\end{document}